\documentclass[aps,prb,superscriptaddress,twocolumn]{revtex4}

\usepackage{graphicx}

\begin{document}

\title{Training of the Ni-Mn-Fe-Ga ferromagnetic shape-memory
alloys by cycling in a high magnetic field}

\author{A.~A.~Cherechukin}

\affiliation{Institute of UHF Semiconductor Electronics of RAS,
Moscow 117105, Russia}

\affiliation{International Laboratory of High Magnetic Fields and
Low Temperatures, Wroclaw, Poland}

\author{V.~V.~Khovailo}

\affiliation{Institute of Fluid Science, Tohoku University, Sendai
980--8577, Japan}

\author{R.~V.~Koposov}

\affiliation{International Laboratory of High Magnetic Fields and
Low Temperatures, Wroclaw, Poland}

\affiliation{Institute of Radioengineering and Electronics of RAS,
Moscow 101999, Russia}

\author{E.~P.~Krasnoperov}

\affiliation{Russian Research Center, Kurchatov Institute, Moscow,
Russia}

\author{T.~Takagi}
\author{J.~Tani}

\affiliation{Institute of Fluid Science, Tohoku University, Sendai
980--8577, Japan}

\begin{abstract}
The temperature and magnetic field dependencies of Ni-Mn-Ga
polycrystals deformation are investigated. Ingots were prepared by
arc-melting in argon atmosphere and further annealing. A training
procedure (cycling across the martensitic transition point) for
the two-way shape-memory effect was performed with
Ni$_{2.16}$Fe$_{0.04}$Mn$_{0.80}$Ga samples. Changes in sample
deformations were noticed with changing the magnetic field at a
constant temperature. The first cycle deformation increment as
compared with the initial value (in the austenitic state at zero
field) in the course of the martensitic transition was 0.29\%, and
0.41\% and 0.48\% for the second and third cycles, respectively.
\end{abstract}



\maketitle

\section{Introduction}

The Heusler alloys of Ni-Mn-Ga system are materials with
well-investigated structural and magnetic properties~\cite{1-w}
that reveal shape-memory effect (SME) as the following process of
the first-order structural phase transition (SPT).
Stress~\cite{2-k} and temperature~\cite{3-c} induced SPT in
Ni-Mn-Ga proceeds in a similar way to that in Ni-Ti where SME was
demonstrated. Alloys with SME are now used as basic elements for
actuators, thermo- and magnetosensors and are widespread in
technique and medicine.

Ferromagnetic order of Ni-Mn-Ga allows one to consider the alloys
to be a prominent kind of "smart materials" that are to be
thoroughly investigated and practically used. Predicted
possibility of magnetic field controlling of the polycrystalline
sample shape was proved to be correct.~\cite{4-u,5-j,6-c,7-c,8-b}
The magnetic field action on deformation of a Ni–Mn–Ga single
crystal was first investigated by Ullakko \textit{et
al.}~\cite{4-u} Associated with twin boundaries motion,
deformations of about 0.2\% had been induced by an external field
of 8~kOe, applied along [001] axis of the crystal. The
field-induced deformation of 4.3\% after mechanical preloading was
reported by James \textit{et al.}~\cite{5-j} In Ref.~6 the one-way
magnetic-field-induced SME was demonstrated. Restorable
deformation value $\epsilon$ reaches 3.5\% at a 100~kOe field. In
this case deformation is caused by SPT.

Magnetic properties of Ni$_{2+x}$Mn$_{1-x}$Ga alloys are
conditioned by Mn atoms at which magnetic moments of 4.17 $\mu_B$
are localized that is strongly more than 0.3 $\mu_B$ at Ni
atoms.~\cite{1-w} Thus partial substitution of Mn by Ni gives a
region of concentrations $x = 0.16 - 0.19$ where magnetic and SPT
are close to each other or coincide.~\cite{8-b} SPT temperatures
of these non-stoichiometric alloys most heavily depend on the
applied magnetic field. Realization of a reversible
magnetic-field-induced SPT at constant temperature and stress for
these concentrations needs the field of about 100~kOe.~\cite{7-c}
Inclusion of Fe allows to enhance the mechanical properties of the
alloy without sacrificing its magnetic and structural properties.

The present work describes an attempt of training a ferromagnetic
Ni$_{2.16}$Fe$_{0.04}$Mn$_{0.80}$Ga polycrystalline sample for
two-way shape memory by magnetic field cycling across the
transition point.

\section{Experimental}

Polycrystalline ingots of Ni$_{2.16}$Fe$_{0.04}$Mn$_{0.80}$Ga were
produced by arc-melting of high-purity initial elements in argon
gas atmosphere on a cold bottom. For homogenisation they were
annealed at 1100~K for 9~days. Then the samples of $8\times
2.5\times 0.36~\mathrm{mm}^3$ were cut out. An optical method was
used to perform the investigation.~\cite{6-c} The maximal
available magnetic field provided by the Bitter magnet was of
85~kOe value. Ordinary weight up to 0.05~kg was used to load the
sample employing the three-point method. The temperature was
varied by an electric oven.

The investigated material has Curie temperature $T_C = 340$~K
higher than the finish reverse martensite transition point $A_f$.
Therefore, it is ferromagnetic both in the austenite and
martensite phases.~\cite{7-c} Martensite transition goes from
martensite transition start temperature $M_s = 312.5$~K to
martensite transition finish $M_f = 305.5$~K for untrained sample.
Reverse transition points are austenite transition start $A_s =
308.5$~K and finish $A_f = 314.0$~K.

The first step for the magnetic field training experiment was
heating the sample to a temperature just above $A_f$. Then the
temperature was stabilized. After this the magnetic field was
slowly being increased from 0 to 85~kOe (for about 3~min) to
induce SPT. After a minute pause, the field was decreased at the
same rate to transfer the sample back to the austenite state. This
was the procedure of a single cycle.

\section{Results and discussion}

The procedure of temperature training has been elaborated. The
temperature dependence of Ni$_{2.16}$Fe$_{0.04}$Mn$_{0.80}$Ga
polycrystal deformation is depicted in Fig.~1. Non-zero initial
deformation is conditioned by the load. For example, the processes
during the first cycle are as follows.

The flat section of the curve from 323 to 314~K is a 100\%
austenite region. At $M_s = 312.5$~K the martensite transition
starts. The crystal lattice transforms from cubic to tetragonal.
This leads to change of volumes occupied by each phase and thus to
considerable deformation of the sample. This section is the
superplasticity region. At heating, one can observe the reverse
process with a temperature hysteresis of about 5~K. As a result
the crystal recovers its initial shape in austenite phase doing
work against the external force. This loop exhibits a one-way SME.
With the number of cycle maximal deformation progressively rises.
This is exactly the training process.

\begin{figure}[t]
\begin{center}
\includegraphics[width=\columnwidth]{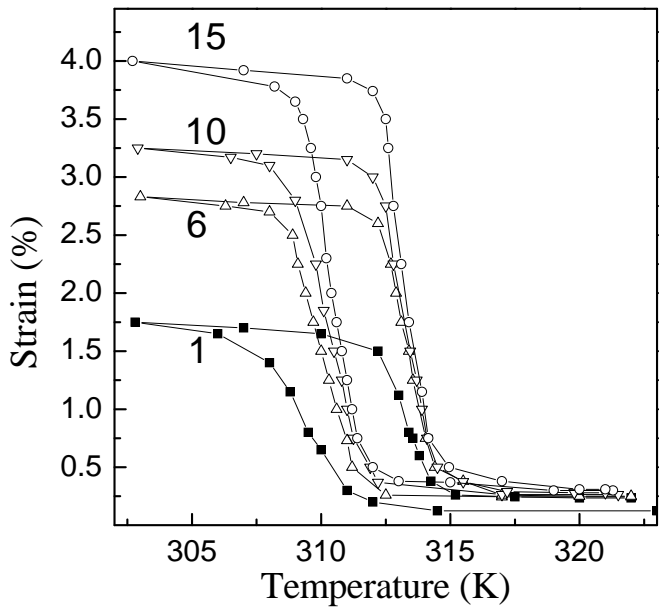}
\caption{$\epsilon - T$ curves of sample loaded by weight of 3~N.
Training by cooling and heating. Cycles 1, 6, 10 and 15 are
shown.}
\end{center}
\end{figure}

The $M_s - M_f$ region shifts to the range of higher temperatures
with the number of processed thermocycle increase. This is
followed by hysteresis loop diminution.~\cite{9-s}

The existence of a reversible magnetic-field-induced SPT at
constant temperature in a magnetic field of about 80 -
100~kOe~\cite{7-c} allows one to change the sample deformation
caused by phase transition not only by temperature change, but
also by magnetic field; and to train the sample for the two-way
SME by this method. The field dependence of deformation for
Ni$_{2.16}$Fe$_{0.04}$Mn$_{0.80}$Ga during training at a constant
temperature of about 314 - 315~K and load $p = 0.5$~N is presented
in Fig.~2. The $\epsilon (H)$ plots at the first, second and third
cycles are shown.

Let us discuss the procedure of field cycling in more detail. As
magnetic field switching off acts as a temperature rise, the
graphs in Figs.~1 and 2 are reflection symmetric; and in this case
we observe all the thermoelastic effects that we do at $\epsilon
(T)$ dependence. These are superplasticity --- actual non-elastic
change of deformation during phase transition caused by
crystalline lattice transformation induced by the magnetic field,
and SME --- the sample shape restoration at field value decrease.
The deformation value at the first cycle is 0.29\% for maximal
martensite, at the second cycle is 0.41\% and at the third cycle
is 0.48\% as compared with the initial value. This increasing of
maximal deformation with the number of cycles indicates the
possibility of attaining the two-way shape memory by magnetic
field training. The incomplete recovering of deformation is caused
by unstable temperature ($\pm$~1~K) and the highest available
field (85~kOe) which was not enough to induce complete reverse
transition.~\cite{7-c}

Hysteresis loops are slightly shifted to lower fields with the
cycle number increase. An effect, related to the martensitic
transition start field decrease (or $M_s$ rise) and hysteresis
loop diminution (in this case from 68~kOe for the first cycle to
60~kOe for the third cycle) is observed like in many materials
with the SME at thermocycling (see, for example, Ref.~9) and can
be explained as follows.

\begin{figure}[h]
\begin{center}
\includegraphics[width=\columnwidth]{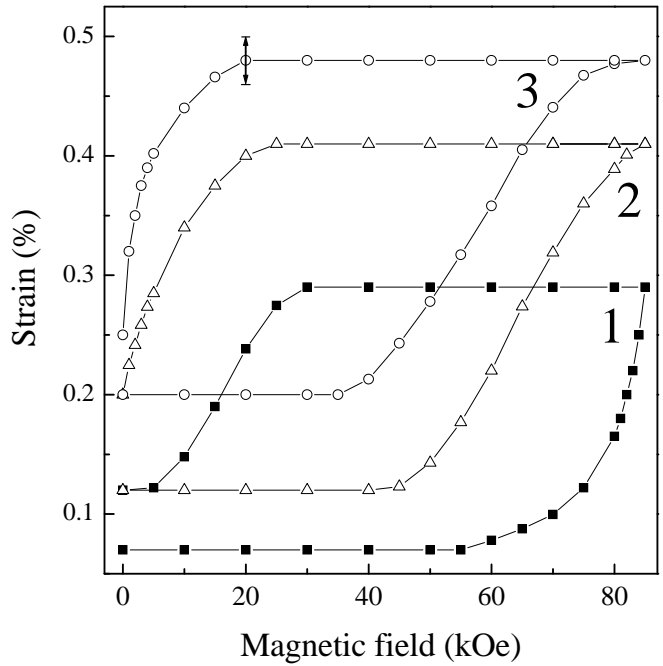}
\caption{$\epsilon - H$ curves of sample loaded by weight of
0.5~N. Training by magnetic field. Cycles 1, 2 and 3 are shown.}
\end{center}
\end{figure}

Vestiges of martensite plates in the form of dislocation tangles
and local stress fields remain in the parent phase after the
repeated formation and reversion of stress-induced martensite. The
dislocation structure developed during cycling assists the
nucleation of the preferred variants of stress-directed
thermoelastic martensite during field rise (just as during
cooling), thus more early martensite transition start and
decreaseing the hysteresis are observed.

\begin{figure}[t]
\begin{center}
\includegraphics[width=\columnwidth]{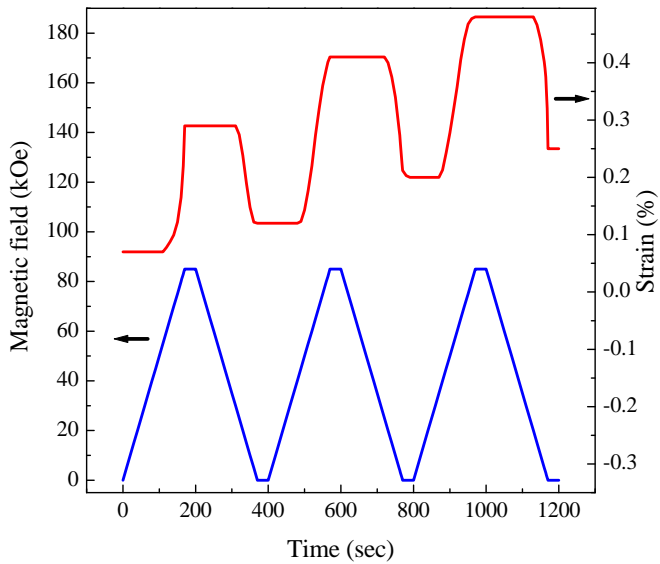}
\caption{Temporal $\epsilon$ and $H$ curves of the sample during
magnetic field cycling.}
\end{center}
\end{figure}

The plot in Fig.~3 is of an illustrative interest. Temporal
dependencies of magnetic field and deformation show some delay of
field-induced deformation change. This is exactly field
hysteresis. The maximal deformation rise with cycle number is
evident while the highest field at each cycle is the same and
equals 85~kOe.

\section{Conclusion}

In conclusion we note that the attempt to train
Ni$_{2.16}$Mn$_{0.80}$Fe$_{0.04}$Ga sample by magnetic field
cycling demonstrated that the material during the cycling revealed
behaviour similar to that during the thermocycling.

There are two reasons related to each other to predict that
magnetic field training of Ni-Mn-Ga alloys permits higher
efficiency of field-induced SPT. The first reason is that the SPT
tends to reverse transition in the same thermodynamic way as the
direct one and this scheme also works at further cycles, the
method of training is a principal factor of field-induced shape
change efficiency. While second reason is that the transition
induced by magnetic field switching off and on structurally
differs from that induced by temperature change. So magnetic field
should influence field cycling trained sample shape more strongly
than the temperature cycling trained one.

\end{document}